# Short–wavelength infrared avalanche photodetector using Sb-based strained layer superlattice

Arash Dehzangi, Jiakai Li, Manijeh Razeghi[1,a)]

[1]Center for Quantum Devices, Department of Electrical Engineering and Computer Science, Northwestern University, Evanston, Illinois 60208

## Abstract

We demonstrate a low noise short–wavelength infrared (SWIR) Sb-based type II superlattice (T2SL) avalanche photodiodes (APD). The SWIR GaSb/(AlAsSb/GaSb) APD structure was designed based on impact ionization engineering and grown by molecular beam epitaxy on GaSb substrate. At room temperature, the device exhibits a 50 % cut-off wavelength of 1.74 µm. The device revealed to have electron dominated avalanching mechanism with a gain value of 48 at room temperature. The electron and hole impact ionization coefficients were calculated and compared to give better pro-spect of the performance of the device. Low excess noise, as characterized by the carrier ionization ratio of ~ 0.07, has been achieved.

Avalanche photodiodes (APDs) internally amplify charge carriers with an avalanche process while operating under a high reverse bias that can cause impact ionization compared to conventional p-i-n photodiodes. APDs can deliver high sensitivity involved with gain mechanism via avalanche multiplication with several applications in military and fiber-optic communication, imaging and commercial sector.[1-5]

For short–wavelength infrared (SWIR) APDs, several material systems are implemented, including silicon, AlGaAs/InGaAsSb, InP/InGaAs, and HgCdTe (MCT) [6-8]. However, the spectral band between 1.5–2.6 µm of the SWIR range can be served further compared to InP/InGaAs or MCT. Lattice matched InGaAs/InP can deliver high performance APD devices operating in the 0.9 to 1.7 µm wavelength range. InGaAs detectors are capable to reach longer cut-off wavelengths by increasing the indium content, however, the crystal defects introduced by the epitaxial process used to extend the indium content degrade the performance as the cut-off wavelength gets longer. MCT on the other hand is the most mature material system for infrared

---

[a)] *Corresponding author:* razeghi@eecs.northwestern.edu

technology, but it suffers from drawbacks due to bulk and surface instability and higher costs particularly for fabrication. [9, 10]

Due to the nature of impact ionization the avalanche process is a random process, which is associated to a factor named excess noise F(M). According to the local-field avalanche theory, both the F(M) and the gain-bandwidth product of an APD can be impacted by the *k* factor which is the ratio of the hole (β) and electron (α) ionization coefficients of the APD. As demonstrated by McIntyre [11] a large difference in the ionization rates for electrons and holes (low *k* factor) is essential for a low noise avalanche photodiode. The F(M) which is given by:

$$F(M) = kM + (1-k)\left(2 - \frac{1}{M}\right) \qquad (1)$$

rises with increasing the gain (M), but the rate of increasing noise can be slow down by reducing the *k* value. Lower *k* value can improve the performance of the APD devices.[12, 13] Therefore low *k* factor is crucial for high-speed and low-noise operation of APD device. The *k* value can be minimized under single-carrier-initiated single-carrier multiplication (SCISCM) conditions (means that an APD must be operated such that only one carrier species ionizes).[14, 15] This is difficult when for some materials the impact ionization coefficients are similar ($\beta/\alpha = k \cong 1$); it is therefore of great interest to explore the possibility of "artificially" decreasing *k* in these materials by using APDs with bandstructure-engineered avalanche regions.

There is a great opportunity for introducing a new material capable of low dark current, high quantum efficiency, and single carrier multiplication for use in strategic SWIR range. Antimony (Sb)-based III/V materials (bulk and superlattice) are capable of meeting the bandgap requirements for making APDs in the SWIR spectral range. In order to achieve great characteristics for Sb-based APD with high gain and low noise, the bandgap and its electron and hole ionization coefficients have to be designed carefully. To minimize the excess noise factor, a pure or dominant electron or hole initiated multiplication along with optimized hetero-junction design can be applied via impact ionization engineering.[16] One of the possible alternative of impact ionization engineering for SWIR APDs is by using the multi-quantum well (MQW) structure as the avalanche region. In MQW-based avalanche region, the impact ionization happens easily between the heterointerfaces between the barrier and well layers due to a sharp bandgap discontinuities.[17]

Sb-based strained layer superlattice (SLS) material is [18] a developing material system with flexible band gap engineering and capabilities to cover the entire range of infrared light using different combination and compositions of Sb based heterostructures, such as InAs/GaSb/AlSb or InAs/InAsSb with Type II staggered gap (type II) band alignment [19-22]. Recently new gain-based structures including APDs based on SLS Sb-base material have also been reported for SWIR region. [23-26]

The flexibility of T2SLs band structure engineering has a significant advantage for designing multi quantum well (MQW)-based APD [27]. In this MQW structure the band discontinuities between well and barrier can be engineered to have a large conduction band discontinuity ($\Delta E_c$) and a small valence band discontinuity ($\Delta E_v$). In the MQW structure, electron ionization rate can be enhanced, since the electrons receive kinetic energy $\Delta E_c$ at hetero interfaces. Holes, on the other hand, can flow unhindered across the MQW because $\Delta E_v$ almost vanishes

In this letter, we demonstrate a SWIR APD structure based on MQW structure consisting of 40 loops of bulk GaSb well layer and AlAsSb/GaSb T2SL structure barrier layer sandwiched between two highly doped contact layers. The schematic of the design and structure of the SWIR APD device is shown in Fig 1a. The device structure was grown on 2-inch Te-doped n-type ($10^{17}$ cm$^{-3}$) GaSb (100) substrate using an Intevac Modular Gen II molecular beam epitaxy (MBE) system. As first step 100 nm thick GaSb buffer layer was grown. Then, a 500 nm thick n-contact ($10^{18}$ cm$^{-3}$) of InAs/GaSb/AlSb/GaSb SLS was grown. Then the absorption layer of 20 loops of undoped MQW consisting of AlAsSb/GaSb barrier layer and GaSb well layer was grown. In such a MQW structure, AlAsSb/GaSb superlattice serves as a barrier layer and GaSb as a well layer. The AlAs$_{0.10}$Sb$_{0.90}$ layer in the AlAsSb/GaSb superlattices is well lattice-matched to GaSb substrate with antimony atom in common with the substrate, this in turn can bring a great range of flexibility in the superlattice growth and design. At last, a 100 nm top p-contact ($10^{18}$ cm$^{-3}$) GaSb layer was grown. During growth, silicon and beryllium were used for n-type and p-type dopant, respectively.

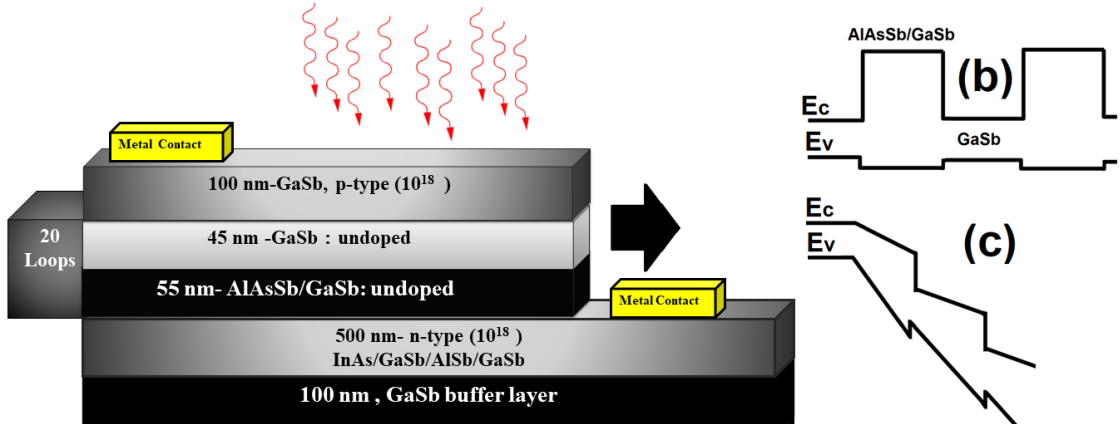

**Figure 1.** (a) Schematic of the SWIR APD structure under top illumination. Energy-band diagram of the MQW structure consisting of AlAsSb/GaSb barrier layer and GaSb well layer (b) unbiased (c) under a bias voltage. The band edge discontinuities are $\Delta E_c \geq 0.5 eV$, $\Delta E_v < 0.15 eV$

The empirical tight-binding method (ETBM) with *sps\** formalism, with nearest neighbor interactions, under a two-center approximation, which was modified from previous work[28] was used to calculate the band discontinuities in absorption region between the AlAsSb/GaSb superlattice barrier (55 nm) and the GaSb (45 nm) well. Both barrier and well region were left undoped. The ETBM material parameter sets in the previous work were used[28]. The $\Delta E_c$ and $\Delta E_v$ between the barrier and well in the MQW structure were calculated to be ~0.50 eV and ~0.15 eV, respectively.

Energy-band diagram of the GaSb/(AlAsSb/GaSb) superlattice structure (unbiased and under bias voltage are schematically illustrated in Fig 1(b), (c). In this MQW structure, consider a hot electron accelerating in an AlAsSb/GaSb barrier layer under the bias voltage applied to the structure. When it enters in the GaSb well, it abruptly gains energy equal to the conduction band offset edge ($\Delta E_c$). The main effect is that the electron goes under a stronger electric field (increased by $\Delta E_c$). In contrast, the hole ionization rate $\beta$ is not substantially increased by the superlattice because the valence-band discontinuity is much smaller, which leads to have a reduction in the *k* value. After the MBE growth, the material quality of the SWIR APD sample was assessed using atomic force microscopy (AFM) and high-resolution X-ray diffraction (HR-XRD).

In order to verify the cut-off wavelength of the SWIR APD devices, they were optically characterized using a temperature and pressure-controlled Janis STVP-100 two chamber liquid helium cryostat station with 300 K background. The optical response of the SWIR APD was done under front–side illumination at room temperature. No anti–reflection coating was applied to the devices. The photodetector spectral response was measured using a Bruker IFS 66v/S Fourier transform infrared spectrometer (FTIR) and the absolute responsivity of the device was calculated using a band-pass filter in front of the calibrated blackbody source at 1000 °C.

The optical performance of the devices is shown in Fig 2. At room temperature, the device exhibits a 50 % cut-off wavelength of 1.74 µm with the responsivity reaches a peak value of 0.38 A/W at 1.65 µm under -5.0 V applied bias, respectively. Fig 2 also shows the photoluminescence of the grown wafer at room temperature with the peak of the spectrum is centered at $\lambda = 1674$ nm.

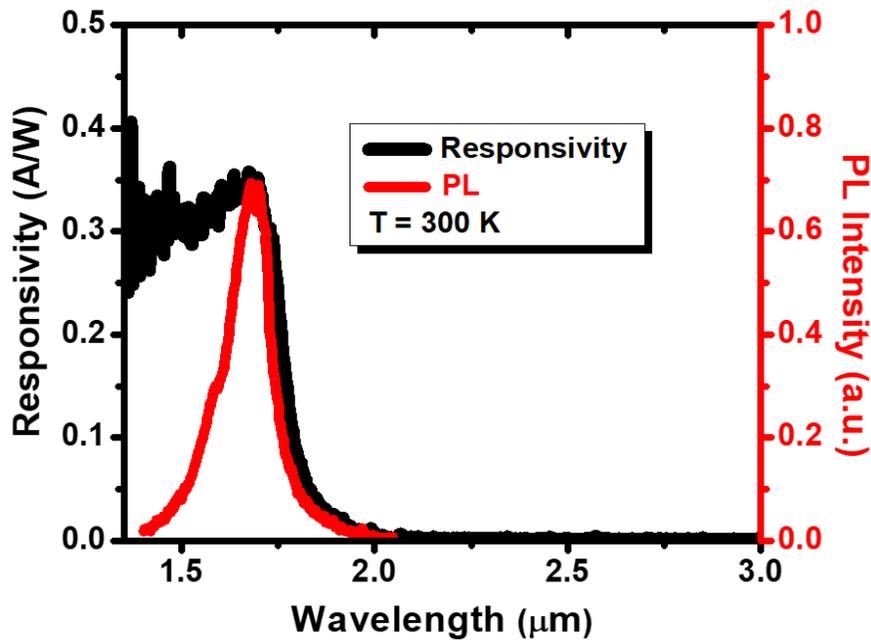

**Figure 2.** Responsivity spectra measured under front-illumination at 300 K under - 5.0 V bias voltage. (Red) The photoluminescence of the wafer at 300K.

Current-voltage (I-V) measurements were carried out using Agilent 4156c semi-conductor parameter analyzer. The temperature of the devices was varied to study the change in gain characteristics at different temperatures. A 633 nm He–Ne laser with an incident power of 5.0 mW was used to measure the photocurrent and the gain of the APDs. Fig 3 shows the I-V characteristics of a 40 µm size SWIR APD with a gain around 48 at a reverse bias voltage of − 50 V at room temperature. The exponential nature of the gain indicates the occurrence of the avalanche mechanism, as seen in similar SWIR APDs. At 300 K, the device shows a unity optical gain dark current of $3.66 \times 10^{-6}$ A at -19 V applied bias. The diodes show punch-through effect at the voltage near $\sim$ - 19 V. This effect is considered to be related to the voltage required to achieve a full depletion of the absorption layer, however, more future study is needed to confirm this hypothesis for present work.

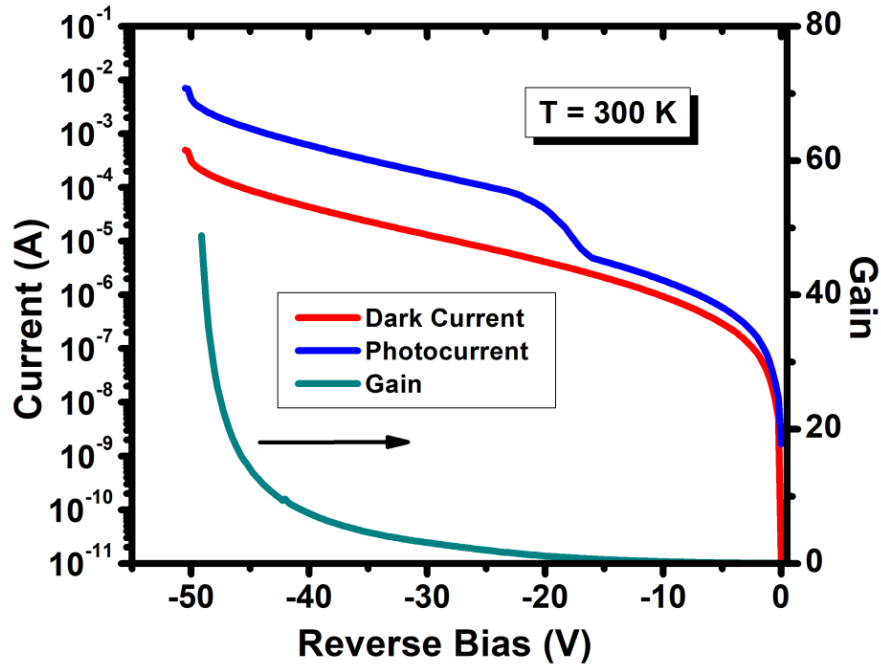

**Figure 3.** Breakdown characteristics and gain of SWIR APD device at 300 K. Dark current and photocurrent are shown on the left axis; gain is shown on the right axis.

We illuminated the device from either top p+ contact and top n+ contact to control the dominate carrier injection into the multiplication region.[29-31] For top contact illumination of n+ contact, a separate device with flipped structure was grown and processed under the same condition. In general, the electron and hole impact ionization coefficients, $\alpha$ and $\beta$ can be derived from the experimental value of electron initiated avalanche gain and hole initiated avalanche gain ($M_e$ and $M_h$) by solving the avalanche rate equations[32]. The extracted electron and hole impact ionization rate for SWIR APD is shown in Fig 5.

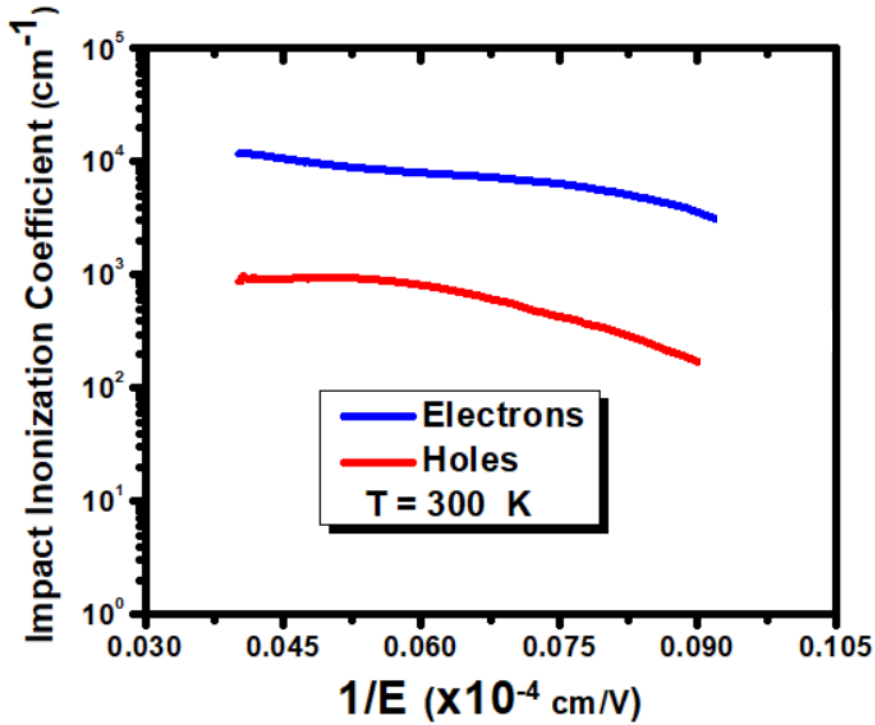

**Figure 4.** Hole and electron ionization coefficients for SWIR APD vs inverse of electric field at T= 300 K

For top contact illumination of n+ contact, a separate device with flipped structure was grown and processed under the same condition. The large difference in α and β (see Figure 4) under these different injection regimes is direct evidence that the effective α is greater than β. It also implies that avalanche multiplication process is dominated by impact ionization of electrons. The impact ionization coefficient for the SWIR APD was calculated to be 0.07 = $k$ at room temperature (300K). This small k ratio is largely due to enhanced electron impact ionization, which also agrees fairly well with theoretical predictions and experimental results of this effect in a superlattice with characteristics similar to ours.[17, 27]

The gain of the device was measured at different temperature as illustrated in Fig 5. At 200 K the gain of SWIR APD device reached around 206 at – 50 V bias voltage and stays constant up to 220 K and then it shows a continuous slow decrease at higher temperatures upon warming up to room temperature. The decrease of the gain value at higher temperature could be associated to a higher probability of carrier-carrier scattering and higher loss of kinetic energy of carriers via various scattering mechanisms, more study is needed to perform in this subject.

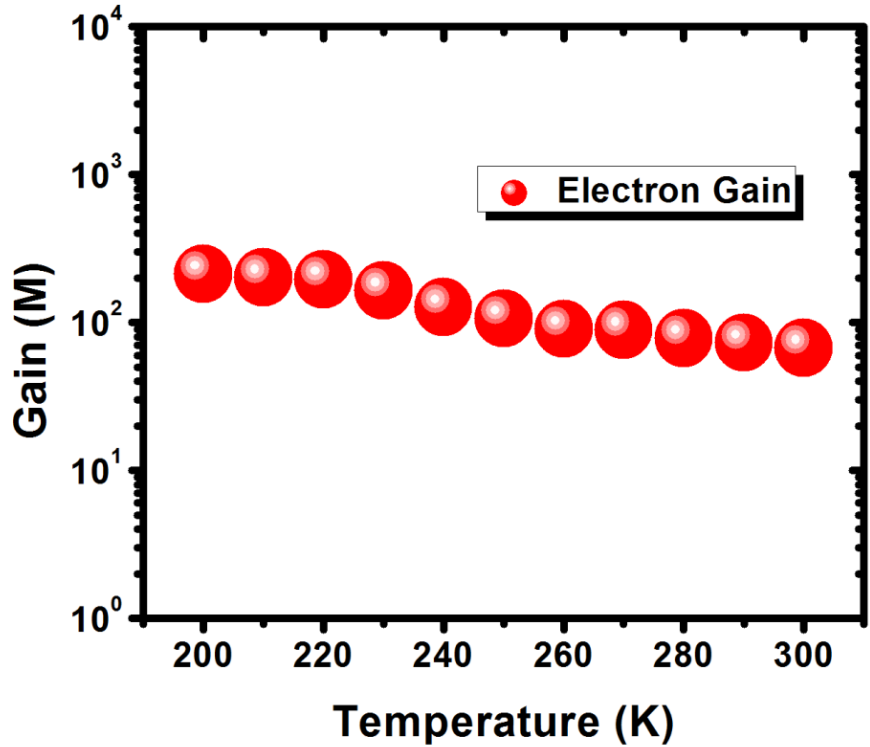

**Figure 5.** Temperature dependent gain characteristics of the SWIR APD from 220 K to 300K at – 50 V bias voltage.

Fig 6 shows the excess noise factors measured for the SWIR APD (using SR770 FFT spectrum analyzer) and comparison with excess noise values predicted by the local field model in Equation 1 [11]. For lower values of multiplication gain at the gain range of 1 to 20, the measured SWIR APD excess noise corresponds to an estimated k-value of 0.07 which was expected, however for higher gain values 20 to 50 there is a small increasing trend as the multiplication gain increases corresponding to k-value between 0.07 and 0.1. This could be due to limitation of measurement system at high biased voltage and yet need to be taken under further study for future direction of research. This k value is close to the best reported values for APD devices based on similar materials, such as AlInAsSb (k= 0.015)[24], AlAsSb (k= 0.1)[33], AlGaAs (k= 0.1)[34] and it is better than k values for commercial Silicon bases APD (k values around 0.02 and 0.06); InP based APDs (k values between 0.4 and 0.5) and InAlAs (k values in the range 0.2 and 0.3).[16, 33, 35] By further development of the device architecture and implementing band engineering in the superlattice system, low noise based SWIR device based on SLS material is achievable for low noise application with higher gain bandwidth product.

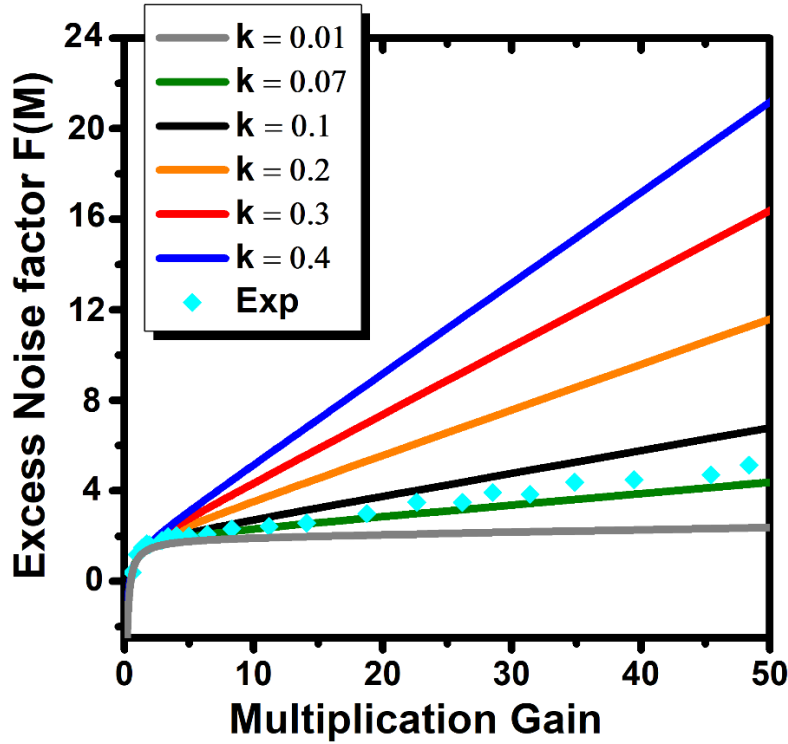

**Figure 6.** Measured excess noise factor at T= 300 K versus gain for the SWIR APD. The solid lines are plots of the excess noise factor using the local field model for k

In summary, using impact ionization engineering Sb-based SLS material structure was implemented to demonstrate a low noise SWIR GaSb/(AlAsSb/GaSb) superlattice APD device. The multiplication gain of 48 was achieved at room temperature. The structure was designed based on impact ionization engineering by implementing the MQW structure. The device exhibits a 50 % cut-off wavelength of 1.74 µm at room temperature. The electron and hole impact ionization coefficients for the SWIR APD device was calculated and compared with each other to give better prospect of the performance. This leads to extracting the carrier ionization ratio with the value of 0.07 for the SWIR APD. The SWIR APDs revealed promising gain/noise characteristics for low noise applications.